Calculation of thermal parameters of SiGe microbolometers

A. V. Voitsekhovskii,[1,2] D. V. Grigoryev,[2] V. A. Yuryev,[3] and S. N. Nesmelov[2]

[1] Tomsk State University;
[2] V. D. Kuznetsov Siberian Physical-Technical Institute at Tomsk State University;
[3] A. M. Prokhorov General Physics Institute of the Russian Academy of Sciences;
e-mail: vav@elefot.tsu.ru.

The thermal parameters of a SiGe microbolometer were calculated using numerical modeling. The calculated thermal conduction and thermal response time are in good agreement with the values found experimentally and range between $2 \cdot 10^{-7}$ and $7 \cdot 10^{-8}$ W/K and 1.5 and 4.5 ms, respectively. High sensitivity of microbolometer is achieved due to optimization of the thermal response time and thermal conduction by fitting the geometry of supporting heat-removing legs or by selection of a suitable material providing boundary thermal resistance higher than $8 \cdot 10^{-3}$ cm$^2 \cdot$K/W at the SiGe interface.

**INTRODUCTION**

Nowadays, infrared imaging devices are an integral part of modern systems applied to solution of a variety of problems of all kinds. From the practical standpoint, the infrared imagers operating in the ranges between 3–5 (MWIR) and 8–12 μm (LWIR) are in most common use due to high atmospheric transmission in the foregoing spectral ranges.

A detector array sensitive to infrared radiation in a certain wavelength range provides a basis for any IR imager. Until recently, the arrays of microbolometers were largely realized on the vanadium oxide VO$_x$ providing high homogeneity, low noise, and reasonably high temperature coefficient of resistance. However, the use of this material leads to several technological problems within the CMOS technology necessitated by providing control of oxygen content in a narrow range.

Polycrystalline or amorphous materials based on Si are most attractive for the purpose. These materials, in spite of a somewhat increased intrinsic noise level as compared with VO$_x$ detectors, provide much lower thermal conduction (up to $8 \cdot 10^{-8}$ W/K). Therefore, much recent attention is given to the development of IR systems on the basis of uncooled bolometric arrays of polycrystalline germanium silicide (SiGe). The application of the detectors in IR systems offers a number of advantages: high quality of infrared images due to high homogeneity of the arrays and a possibility to produce large arrays, relatively low cost of the arrays, and a possibility of integrating the process of producing arrays into a conventional silicon technology, *etc*.

To develop effective photodetectors, it is necessary that these possess optimal operating parameters (signal, noise, and threshold ones). To determine the operating characteristics of a detector, it is necessary to calculate the characteristics of an individual microbolometer element, these being dependent on its thermal parameters.

The aim of the work is to calculate the thermal characteristics of a SiGe microbolometer, which describe its temperature history under external actions.

**GENERAL FORMULATION OF THE PROBLEM**

The thermal characteristics of a microbolometer such as the thermal response time $\tau_{th}$ and thermal conduction $G_{th}$ were calculated using numerical modeling of the time history of the temperature field $T(x, y)$ occurring in the microbolometer exposed to a short thermal pulse causing instantaneous heating of a sensitive plate by one degree. The time history was calculated by means of solving the thermal conduction equation with allowance for a particular microbolometer structure.

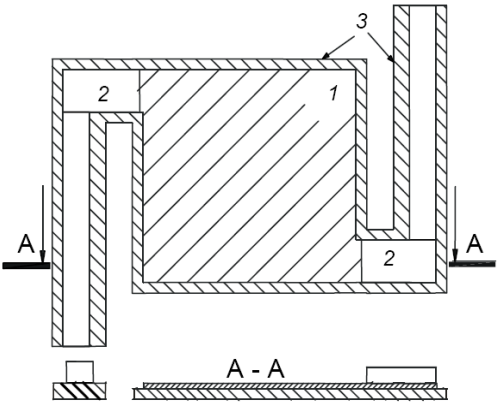

Fig. 1. Microbolometer structure: sensitive plate SiGe (*1*), heat-removing legs (doped SiGe, aliminium) (*2*), and substrate ($Si_3N_4$) (*3*).

The integral temperature of the microbolometer sensitive plate $T_d$ was found by calculating the integral

$$T_d = \frac{1}{A_d} \iint_{A_d} T(x,y)\,dxdy\,,$$

where $A_d$ is the microbolometer sensitive-plate area.

The temperature relaxation of the bolometer sensitive surface is described by the following relation:

$$\frac{d\Delta T}{dt} + \frac{\Delta T}{\tau_{th}} = 0\,,$$

where $\tau_{th}$ is the microbolometer thermal response time. The value of $\tau_{th}$ was calculated using approximation of the time history of the integral temperature of the bolometer sensitive plate by exponential dependence. The bolometer thermal conduction $G_{th}$ due to heat-removing legs was calculated from the derived value $\tau_{th}$ as follows:

$$G_{th} = \frac{C_{th}}{\tau_{th}}\,,\quad C_{th} = A_d \rho c_d t_d\,,$$

where $C_{th}$ is the specific heat, $t_d$ is the microbolometer sensitive-plate thickness, and $\rho$ and $c_d$ are the density and specific heat of the material of the sensitive plate.

The structure of the microbolometer (Fig. 1), for which the thermal parameters were calculated, was determined from analysis of the literature data [1–4]. The microbolometer consists of a sensitive plate made from polycrystalline SiGe and supports playing the role of heat-removing legs and electrical contacts. To provide an electrical contact, the supports are made either from highly doped SiGe or aluminium. The sensitive plate and supports are on a $Si_3N_4$ substrate. Since $Si_3N_4$ exhibits high thermal resistance ($10^8$ K/W [5]), its effect can be neglected in the mathematical description of the problem.

In the general case, a three-dimensional heat conduction equation must be solved for the microbolometer structure under study. In numerical calculations, we assume that the thermal equilibrium over the structure thickness is established for the time much less than a time increment. In this case, the problem can be reduced to a two-dimensional one. To verify the assumption, we need to solve the problem on heat distribution over the structure thickness, where the temperature is kept constant at one boundary ($z = 0$), whereas there is no heat flow at the second boundary ($z = L$). That is, one must solve the following heat conduction equation:

$$\frac{\partial T}{\partial t} = a^2 \frac{\partial^2 T}{\partial z^2}\,,\quad T(z,0) = T_0\,,\quad T(0,t) = T_G\,,\quad \left.\frac{\partial T}{\partial z}\right|_{z=L} = 0\,. \tag{1}$$

Here $T(z, t)$ is the desired temperature distribution, $L$ is the structure thickness, $T_0$ and $T_G$ are the initial temperature of the structure and that at the boundary, respectively, and $a^2 = g_d/(c_d \cdot \rho)$, where $g_d$, $c_d$, $\rho$ are the specific heat, conduction, and density of the material, respectively.

Solving the foregoing equation by the method of separation of variables [6], we derive the temperature distribution over the structure thickness as

$$T(z,t) = T_G - \sum_{n=1}^{\infty}\left[(-1)^{n+1}\frac{4(T_G - T_0)}{\pi(2n-1)}\sin\left(\frac{\pi(2n-1)}{2L}z\right)\exp\left(-a^2\left(\frac{\pi(2n-1)}{2L}\right)^2 t\right)\right].$$

(2)

The temperature changes over the bolometer thickness and heat-removing legs are plotted in Fig. 2. It is seen that the time of establishing thermal equilibrium over the sensitive plate and heat-removing leg is ~$5 \cdot 10^{-9}$ s. This result allows us to believe that the thermal problem for the bolometer can be solved as a two-dimensional one at a time increment of about 5 ns (Fig. 1).

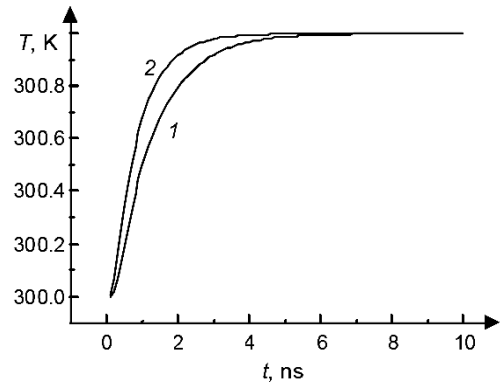

Fig. 2. Temperature distribution $T$ over an 0.1 μm bolometer sensitive plate (*1*) and an 0.4 μm aluminium heat-removing leg (*2*) as a function of time $t$.

To perform simulations, the complex microbolometer structure was presented as a number of ordinary rectangular regions, and a two-dimensional conduction equation with the corresponding boundary conditions was solved for each region. The foregoing equation is written in the general form as

$$c_i\rho_i\frac{\partial T_i}{\partial t} = g_i\left(\frac{\partial^2 T_i}{\partial x^2} + \frac{\partial^2 T_i}{\partial y^2}\right) + Q_i,$$

(3)

$$\left.\frac{\partial T}{\partial x(y)}\right|_{G_{i<>j}} = 0, \quad \left.g_i\frac{\partial T}{\partial x(x)}\right|_{G_{i=j}} = \left.g_j\frac{\partial T}{\partial x(y)}\right|_{G_{i=j}}.$$

(4)

Here $\rho_i$, $c_i$, and $g_i$ are the density, specific heat, and specific heat conduction of the material of the $i_{th}$ region, $T_i \equiv T_i(x, y)$ is the two-dimensional temperature field, $Q_i \equiv Q_i(x, y)$ is the function describing the distribution density of heat sources, $G_{i \diamond j}$ is the boundary of the region $i$, where there is no contact with any neighboring region $j$, $G_{i = j}$ is the boundary of the region $i$, where it contacts any neighboring region $j$, and $g_i$ and $g_j$ are the specific heat conductions of the neighboring regions $i$ and $j$. The boundary conditions (4) are written for the case where the heat flow is zero at the boundary of the regions where there is no contact with the neighboring region.

It should be noted that for such formulation of the problem no account is taken of

1) static stresses occurring at a specified pixel thickness (0.1 μm) which can somehow change thermal parameters of the bolometer and

2) heat sources due to flowing bias current including pulsed application of the bias current.

In addition, it was assumed that the bolometer temperature gradients due to received radiation are fairly low for additional dynamic stresses to emerge. Under such formulation of the problem, its solution allows us to determine the thermal response time $\tau_{th}$ and heat conduction $G_{th}$ due to heat removal by microbolometer legs alone.

In the case, where model calculations of thermal parameters were performed for a microbolometer whose heat removing legs were made from aluminium, additional terms describing heat exchange at the SiGe/Al interface should be introduced into the mathematical model.

The heat exchange at the SiGe/Al interface was taken into account by introducing an additional term describing heat exchange between contacting media into the heat source function, and we have

$$Q_{ij}^G = \frac{B(T_i^4 - T_j^4)}{t_d}, \tag{5}$$

where $t_d$ is the microbolometer thickness, $T_i$ and $T_j$ are the temperatures at the boundaries in the regions $i$ and $j$, respectively.

According to the model of acoustic mismatch [7], the heat flow through the interface of two media is written as

$$W = B(T_1^4 - T_2^4) \tag{6}$$

Here $T_1$ and $T_2$ are the temperatures at the phase boundary of media 1 and 2, respectively and $B$ is the constant depending on the properties of the contacting media (density and velocity of sound). For low heat flows this relation is linearized as

$$W = (T_1 - T_2)/R_{bd},  \tag{7}$$

where $R_{bd} = (4BT^3)^{-1}$ is the boundary heat resistance.

The value of $R_{bd}$ is calculated according to the procedure developed in [8], where the boundary heat resistance is calculated within the model of diffusive phonon mismatch [7]. In this case, $R_{bd}$ can be found as follows:

$$R_{bd} = \left[ \frac{1}{2} \sum_j \upsilon_{1,j} \Gamma_{1,j} \int_0^{\omega_1^D} \hbar\omega \frac{dN_{1,j}(\omega, T)}{dT} d\omega \right]^{-1},  \tag{8}$$

where

$$N_{1,j}(\omega, T) = \frac{\omega^2}{2\pi^2 \upsilon_{1,j}^3 \left[ \exp\left(\frac{\hbar\omega}{k_B T}\right) - 1 \right]}, \quad \Gamma_{1,j} = \frac{1}{2} \frac{\sum_j \upsilon_{2,j}^{-2}}{\sum_{i,j} \upsilon_{i,j}^{-2}}, \quad \upsilon_S = \frac{1}{2} \sum_j \upsilon_{1,j}^{-2}, \quad \omega_1^D = \upsilon_S k_d.  \tag{9}$$

Here $k_B$ is the Boltzmann constant, $\omega$ is the phonon frequency, $T$ is the temperature, $\hbar$ is the Planck constant, $\Gamma_{1,j}$ is the mean coefficient of acoustic wave propagation, $\upsilon_{i,j}$ are the phonon velocities (subscript $i$ corresponds to the medium number and subscript $j = 1$ and $2$ determines longitudinal and transverse sound velocities, respectively), and $\omega_1^D$ is the Debye frequency. The boundary wave vector magnitude is found as

$$k_d = \left( 6\pi^2 N_A \frac{\rho}{M} \right)^{\frac{1}{3}},  \tag{10}$$

where $N_A$ is the Avogadro number, $\rho$ is the density, and $M$ is the atomic weight of the material of medium 1.

It should be noted that there are no data on the velocity of sound in polycrystalline SiGe in the literature, therefore the velocity of sound in polycrystalline SiGe was approximated by the velocity of sound in single-crystal SiGe calculated using the corresponding elastic constants [9] as follows:

$$\upsilon_\perp = \sqrt{\frac{C_{44}}{\rho}}, \quad \upsilon_\parallel = \sqrt{\frac{C_{11}}{\rho}}.  \tag{11}$$

The set of equations (3) – (4) for three connected regions was solved by the explicit finite difference scheme [10]. The parameters of the materials used in the calculations are given in Table 1.

TABLE 1. SiGe and Al Parameters Used in Calculations

| Parameter | SiGe [9] | Al [11] |
|---|---|---|
| $k$, W·cm$^{-1}$·K$^{-1}$ | 0.0712 | 2.09 |
| $c$, J·g$^{-1}$·K$^{-1}$ | 0.583 | 0.88 |
| $\rho$, g·cm$^{-3}$ | 3.33 | 2.7 |
| $M$, a.m.u. | 41.44 | 26.98 |
| $\upsilon_\perp$, cm·s$^{-1}$ | $4.8 \cdot 10^4$ | $3.04 \cdot 10^5$ |
| $\upsilon_\parallel$, cm·s$^{-1}$ | $6.8 \cdot 10^4$ | $6.24 \cdot 10^5$ |

**RESULTS OF MODEL CALCULATIONS AND DISCUSSION**

The thermal parameters were calculated for two microbolometer constructions. In the first construction, the supporting heat-removing legs were made from doped SiGe, and in the second construction – from aluminium.

The calculated temperature distribution over the surface of sensitive plate for the first construction is shown in Fig. 3. The temperature profiles are given for the time of a significant temperature gradient. It is seen from Fig. 3 that the calculation qualitatively describes the problem under consideration. Due to heat removal through the legs to the thermostat, there is a continuous decrease in the temperature of sensitive plate.

The dynamics of integral temperature for the bolometer is plotted in Fig. 4. For a heat removing leg width of 8 μm, the thermal response time is $\tau_{th}$ = 1.5 ms, which corresponds to the heat conduction $G_{th} = 2 \cdot 10^{-7}$ W/K.

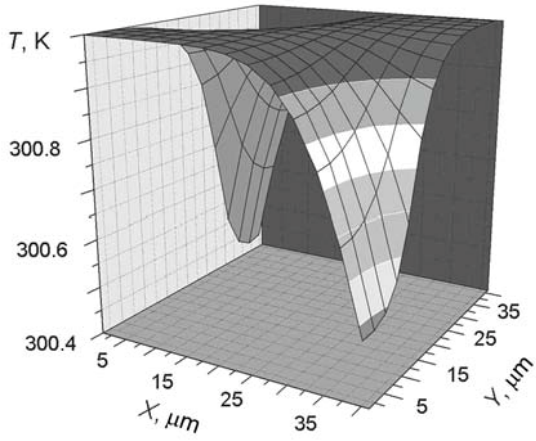 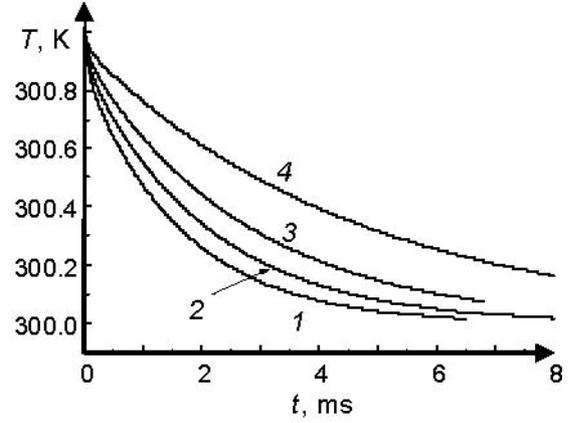

Fig. 3              Fig. 4

Fig. 3. Calculated integral temperature distribution over bolometer sensitive-element surface.

Fig. 4. Dynamics of integral temperature for a microbolometer with SiGe heat-removing supports. The beam width $d$, μm: 8 (*1*), 6 (*2*), 4 (*3*), and 2 (*4*).

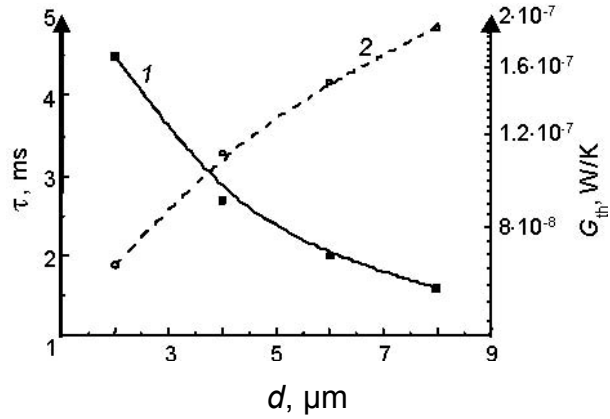

Fig. 5. Microbolometer thermal response time $\tau_{th}$ (*1*) and thermal conduction $G_{th}$ (*2*) versus heat-removing leg width $d$.

The bolometer detectivity is determined by the following relation [5]:

$$D^* = \frac{K\varepsilon R_{th} A^{1/2}}{\left(1+\omega^2\tau_{th}^2\right)^{1/2}\left(\dfrac{4k_B T_d^2 K^2 R_{th}}{1+\omega^2\tau_{th}^2} + 4k_B T_d R + V_{1/f}^2\right)}, \quad (12)$$

where ε is the detectivity of the bolometer, $A$ and $T_d$ are the area and temperature of the bolometer sensitive plate, $R$ and $R_{th}$ are the bolometer electrical and heat resistance ($R_{th} = 1/G_{th}$), $V_{1/f}$ is the noise stress of the type $1/f$, $\omega$ is the modulation frequency of a received signal, and $K$ is the conversion factor of temperature changes in the sensitive plate into output voltage.

It is clear from Eq. (12) that it is necessary to achieve as low heat conduction and as high heat response time as possible to provide high sensitivity of the bolometer. A decrease in the heat conduction of the microbolometer and an increase in its heat response time can be achieved by decreasing the widths of supporting heat-removing legs. Thus, for the width about 2 μm, the microbolometer heat conduction is $G_{th} \sim 7 \cdot 10^{-8}$ W/K, with the heat response time being increased up to $\tau_{th} = 4.5$ ms (Fig. 5).

The experimental heat conduction values for such microbolometer structures range between $10^{-6}$ and $10^{-8}$ W/K. In this case, the heat response time is varied between 1.5 and 20 ms. The calculated thermal parameters allow us to conclude that the developed method is adequate for calculating thermal characteristics of microbolometers.

The calculations performed for the microbolometer structure, where the heat-removing leg was made from aluminium, shows that in this case, a more drastic decrease in the sensitive-plate temperature is observed. This was accompanied by a more intensive heat removal by the supporting legs. For the leg width 8 μm, the microbolometer heat conduction is increased up to $G_{th} = 10^{-6}$ W/K, with the heat response time being decreased down to 0.28 ms. The increase in the microbolometer heat conduction will result in decreasing its sensitivity.

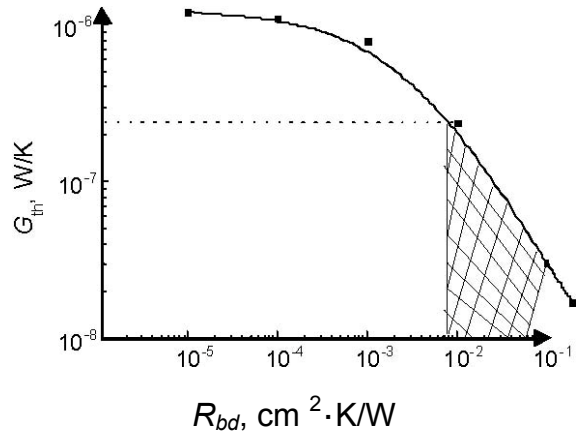

Fig. 6. Microbolometer thermal conduction $G_{th}$ versus boundary resistance between the materials of bolometer sensitive plate and heat-removing legs. The leg width $d = 8$ μm.

In our thermal microbolometer model (3) – (5), heat removal is determined by the heat exchange at the SiGe/Al phase boundary, that is, by the magnitude of thermal boundary resistance $R_{bd}$. The results of investigation into the effect of $R_{bd}$ on the bolometer thermal characteristics are shown in Fig. 6. The magnitude of $R_{bd}$ was varied within three orders of the value $R_{bd} = 2.7 \cdot 10^{-3}$ cm$^2 \cdot$K/W theoretically calculated using Eqs. (8) – (11). It follows from Fig. 6 that the change in $R_{bd}$ by four orders of magnitude results in the change of thermal conduction by a factor of 15. The dashed line in the figure characterizes the thermal conduction of the microbolometer whose heat-removing legs are made from SiGe ($G_{th} \sim 7 \cdot 10^{-8}$ W/K). The shaded region determines the range of boundary resistance, a material of heat-removing legs should possess, for the microbolometer thermal conduction to be lower than its conduction in the case of SiGe legs. Thus, in the case where niobium is used as a material for a heat-removing leg ($R_{bd}$ characterizing the SiGe/Nb interface [12] is 0.2 cm$^2 \cdot$K/W), the microbolometer thermal conduction is $\sim 3 \cdot 10^{-8}$ W/K, with the corresponding thermal response time being 10 ms. A further decrease in $G_{th}$ allows us to achieve a limit on the microbolometer sensitivity by limiting photon fluctuations.

**CONCLUSION**

Our calculations of thermal parameters of a SiGe microbolometer show that the calculated values of thermal conduction and thermal response time are in good agreement with the experimental values. In the case, where the heat-removing legs are made from SiGe, the thermal conduction and thermal response time of the microbolometer range between $2 \cdot 10^{-7}$ and $7 \cdot 10^{-8}$ W/K and between 1.5 and 4.5 ms for the leg width of 2 – 8 μm.

In the case, where the legs are made from aluminium, the calculated value of thermal conduction is increased up to $G_{th} = 10^{-6}$ W/K, however, taking into account experimental boundary resistance (SiGe/Nb), the thermal conduction magnitude is decreased down to $3 \cdot 10^{-8}$ W/K.

The microbolometer thermal response time and thermal conduction can be optimized by selecting either the geometry of the heat-removing leg or its material providing boundary thermal resistance at the phase boundary with SiGe higher than $8 \cdot 10^{-3}$ cm$^2 \cdot$K/W.